\documentclass[a4paper,english,runningheads]{llncs}
\usepackage[T1]{fontenc}
\usepackage[latin9]{inputenc}
\usepackage{amsmath}
\usepackage{amssymb}

\makeatletter

\usepackage[all]{xy}

\usepackage{cite}

\usepackage{babel}

\usepackage{babel}

\makeatother

\usepackage{babel}
\begin{document}

\title{Measuring Observable Quantum Contextuality}

\titlerunning{Measuring Contextuality}

\author{J. Acacio de Barros\textsuperscript{1}, Ehtibar N. Dzhafarov\textsuperscript{2},
Janne V. Kujala\textsuperscript{3}, Gary Oas\textsuperscript{4}}

\authorrunning{J.A. de Barros, E.N. Dzhafarov, J.V. Kujala, G. Oas}

\institute{\textsuperscript{1}San Francisco State University\\
barros@sfsu.edu\\
 $\,$\\
\textsuperscript{2}Purdue University\\
 ehtibar@purdue.edu\\
 $\,$\\
 \textsuperscript{3}University of Jyväskylä\\
 jvk@iki.fi\\
 $\,$\\
 \textsuperscript{3}Stanford University\\
 oas@stanford.edu}

\toctitle{Measuring observable quantum contextuality}

\tocauthor{de Barros, Dzhafarov, Kujala, Oas}
\maketitle
\begin{abstract}
Contextuality is a central property in comparative analysis of classical,
quantum, and supercorrelated systems. We examine and compare two well-motivated
approaches to contextuality. One approach (``contextuality-by-default'')
is based on the idea that one and the same physical property measured
under different conditions (contexts) is represented by different
random variables. The other approach is based on the idea that while
a physical property is represented by a single random variable irrespective
of its context, the joint distributions of the random variables describing
the system can involve negative (quasi-)probabilities. We show that
in the Leggett-Garg and EPR-Bell systems, the two measures essentially
coincide. 
\end{abstract}

\section{Introduction}

Contextuality is a key feature of quantum systems, as no noncontextual
hidden-variable theory exists that is consistent with quantum theory.
This feature has been at the core of recent research in quantum information,
such as attempts to identify the underlying principles for the quantum
boundary. Despite its importance, there seem to be no universally
accepted measure of contextuality, and it is clear that the many definitions
proposed in the literature \cite{larsson_kochen-specker_2002,kleinmann_memory_2011,kurzynski_entropic_2012,chaves_entropic_2012,svozil_how_2012,grudka_quantifying_2014,dzhafarov_contextuality_2014,dzhafarov_probabilistic_2014,dzhafarov_all-possible-couplings_2013,dzhafarov_random_2013,dzhafarov_generalizing_2014,dzhafarov_qualified_2014,DKL2014,KDL2014}
are not all equivalent. Here, we consider and compare two measures
inspired by the idea that contextuality means the impossibility of
finding a joint probability distribution (jpd) for different sets
of random variables with some elements in common. One measure (related
in various ways to Refs. \cite{larsson_kochen-specker_2002,dzhafarov_all-possible-couplings_2013,dzhafarov_qualified_2014,Simon-Brukner-Zeilinger,Winter2014,svozil_how_2012,DK_PLOS_2014}
and in its current form presented in Refs. \cite{DKL2014,KDL2014,KDconjecture,DKL_overview,dzhafarov_contextuality_2014,dzhafarov_generalizing_2014,dzhafarov_probabilistic_2014})
is based on extended sets of context-indexed random variables and
another on negative (quasi-)probabilities (dating back to Dirac, and
recently explored in connection to contextuality in Refs. \cite{spekkens_negativity_2008,abramsky_sheaf-theoretic_2011,de_barros_decision_2014,de_barros_negative_2015,oas_exploring_2014,de_barros_negative_2014}).

As an example of contextuality, let there be three properties of a
system, $P$, $Q$, and $R$, whose measurement outcomes are represented
by the random variables $\mathbf{P}$, $\mathbf{Q}$, and $\mathbf{R}$%
\footnote{This example has the same structure as the Leggett-Garg system \cite{leggett_quantum_1985}.
A special version of it was examined by Suppes and Zanotti \cite{suppes_when_1981},
and also by Specker \cite{specker_logic_1975}.%
}. Assume we can never observe $P$, $Q$, and $R$ simultaneously,
but only in pairwise combinations, $\left(\mathbf{P},\mathbf{Q}\right)$,
$\left(\mathbf{P},\mathbf{R}\right)$, or $\left(\mathbf{Q},\mathbf{R}\right)$.
We may think of each pair as recorded under a different experimental
condition providing a context. The system exhibits contextuality if
one cannot find a jpd of $\left(\mathbf{P},\mathbf{Q},\mathbf{R}\right)$
that agrees with the observed distributions of $\left(\mathbf{P},\mathbf{Q}\right)$,
$\left(\mathbf{P},\mathbf{R}\right)$, and $\left(\mathbf{Q},\mathbf{R}\right)$
as its marginals. The two approaches to be considered in this paper
deal with this situation differently. The negative probabilities (NP)
approach relaxes the notion of a jpd by allowing some (unobservable)
joint probabilities for $\left(\mathbf{P},\mathbf{Q},\mathbf{R}\right)$
to be negative. The ``contextuality-by-default'' (CbD) approach
treats random variables recorded under different conditions as different
``by default'', so that, e.g., property $P$ in the context of experiment
$\left(\mathbf{P},\mathbf{Q}\right)$ is represented by some random
variable $\mathbf{P}_{A}$, and in the context $\left(\mathbf{P},\mathbf{R}\right)$
by another random variable, $\mathbf{P}_{B}$. Denoting the three
contexts by $A,B,C$, this yields three pairs of contextually labeled
random variables, $\left(\mathbf{P}_{A},\mathbf{Q}_{A}\right)$, $\left(\mathbf{P}_{B},\mathbf{R}_{B}\right)$,
and $\left(\mathbf{Q}_{C},\mathbf{R}_{C}\right)$, and in the CbD
approach the joint distribution imposed on them allows, say, $\mathbf{P}_{A}$
and $\mathbf{P}_{B}$ to be unequal with some probability.

Here we compare the NP and CdB approaches applied to the simplest
contextual case possible, given by three pairwise correlated random
variables, and to the standard EPR-Bell experiment. We show that for
such examples the two measures of contextuality are the same.

\section{Negative Probabilities (NP) }

Using our above example, with $\mathbf{P},\mathbf{Q},\mathbf{R}$
observed in pairs, in the NP approach one ascribes to the vector $\left(\mathbf{P},\mathbf{Q},\mathbf{R}\right)$
a joint quasi-distribution by means of assigning to each possible
combination $w=\left(p,q,r\right)$ a real number $\mu\left(w\right)$
(possibly negative), such that 
\begin{equation}
\begin{array}{c}
\sum_{r}\mu\left(w\right)=\Pr\left[\mathbf{P}=p,\mathbf{Q}=q\right],\\
\sum_{q}\mu\left(w\right)=\Pr\left[\mathbf{P}=p,\mathbf{R}=r\right],\\
\sum_{p}\mu\left(w\right)=\Pr\left[\mathbf{Q}=q,\mathbf{R}=r\right].
\end{array}\label{eq:NP exmaple}
\end{equation}
Such $\mu$ exists if and only if the no-signaling condition (built
into EPR paradigms with spacelike separation) is satisfied \cite{al-safi_simulating_2013,oas_exploring_2014,abramsky_operational_2014},
i.e., the distribution of, say, $\mathbf{P}$ is the same in $\left(\mathbf{P},\mathbf{Q}\right)$
and in $\left(\mathbf{P},\mathbf{R}\right)$.%
\footnote{No-signaling condition is a fundamental limitation of any approach
with noncontextually labeled random variables, including NP.%
} The numbers $\mu\left(w\right)$ can then be interpreted as quasi-probabilities
of events $\left\{ w\right\} $, with the quasi-probability of any
other event (subset of $w$ values) being computed by additivity,
inducing thereby a signed measure \cite{halmos_measure_1974} on the
set of all events. The quasi-probability of the entire set of $w$
will then be necessarily equal to unity, because, e.g., 
\begin{equation}
1=\sum_{p,q}\Pr\left[\mathbf{P}=p,\mathbf{Q}=q\right]=\sum_{w}\mu\left(w\right).
\end{equation}
The function $\mu$ is generally not unique. In our approach \cite{de_barros_decision_2014,de_barros_negative_2015}
we restrict the class of possible $\mu$ to those as close as possible
to a proper jpd by requiring that the L1 norm of the probability distribution,
defined by $M=\sum_{w}\left|\mu\left(w\right)\right|,$ be minimized.
This ensures that if the class of all possible $\mu$ satisfying (\ref{eq:NP exmaple})
contains proper probability distributions, the chosen $\mu$ will
have to be one of them. Since in this case $\left|\mu\left(w\right)\right|=\mu\left(w\right)$
for all $w$, the minimum of $M$ is 1. If (and only if) no proper
probability distribution exists, then the minimum of $M$ exceeds
1. As a result, the smallest possible value $\Gamma_{\min}$ of $M-1$
can be taken as a measure of contextuality.

\section{Contextuality-by-Default (CbD)}

A more direct approach to contextuality \cite{dzhafarov_all-possible-couplings_2013,dzhafarov_random_2013,dzhafarov_contextuality_2014,dzhafarov_generalizing_2014,dzhafarov_probabilistic_2014,dzhafarov_qualified_2014,KDL2014,DKL2014,DK_PLOS_2014,DKL_overview,KDconjecture}
is to posit that the identity of a random variable is determined by
all conditions under which it is recorded. Thus, in $\left(\mathbf{P}_{A},\mathbf{Q}_{A}\right)$,
$\left(\mathbf{P}_{B},\mathbf{R}_{B}\right)$, and $\left(\mathbf{Q}_{C},\mathbf{R}_{C}\right)$
of our example, any random variable in any of the pairs is \emph{a
priori }different from and stochastically unrelated to any random
variable in any other pair \cite{dzhafarov_contextuality_2014,dzhafarov_qualified_2014},
but a jpd can always be imposed on the six random variables. In other
words, one can always assign probability masses $\lambda$ to $v=\left(p_{A},p_{B},q_{A},q_{C},r_{B},r_{C}\right)$
in such a way that 
\begin{equation}
\begin{array}{c}
\sum_{p_{B},q_{C},r_{B},r_{C}}\lambda\left(v\right)=\Pr\left[\mathbf{P}_{A}=p_{A},\mathbf{Q}_{A}=q_{A}\right],\\
\sum_{p_{A},q_{A},q_{C},r_{C}}\lambda\left(v\right)=\Pr\left[\mathbf{P}_{B}=p_{B},\mathbf{R}_{B}=r_{B}\right],\\
\sum_{p_{A},p_{B},q_{A},r_{B}}\lambda\left(v\right)=\Pr\left[\mathbf{Q}_{C}=q_{C},\mathbf{R}_{C}=r_{C}\right].
\end{array}\label{eq:CbD exmaple}
\end{equation}
The noncontextuality hypothesis for $\mathbf{P}_{A},\mathbf{Q}_{A},\mathbf{R}_{B}$
and $\mathbf{P}_{B},\mathbf{Q}_{C},\mathbf{R}_{C}$ is that among
these jpds $\lambda$ we can find at least one for which $\Pr\left[\mathbf{P}_{A}\not=\mathbf{P}_{B}\right]=\Pr\left[\mathbf{Q}_{A}\not=\mathbf{Q}_{C}\right]=\Pr\left[\mathbf{R}_{B}\not=\mathbf{R}_{C}\right]=0,$
which is equivalent to $\Delta=\Pr\left[\mathbf{P}_{A}\not=\mathbf{P}_{B}\right]+\Pr\left[\mathbf{Q}_{A}\not=\mathbf{Q}_{C}\right]+\Pr\left[\mathbf{R}_{B}\not=\mathbf{R}_{C}\right]=0.$
Such a jpd need not exist, and then the smallest possible value $\Delta_{\min}$
of $\Delta$ for which a jpd of $\left(\mathbf{P}_{A},\mathbf{P}_{B},\mathbf{Q}_{A},\mathbf{Q}_{C},\mathbf{R}_{B},\mathbf{R}_{C}\right)$
exists can be taken as a measure of contextuality.%
\footnote{This formulation is predicated on no-signaling, which we assume throughout
this paper. CbD has been generalized to situations when this condition
is violated \cite{dzhafarov_probabilistic_2014,bacciagaluppi_leggett-garg_2014,dzhafarov_generalizing_2014,KDL2014,DKL2014,KDconjecture,DKL_overview}.%
} 

The CdB approach has its precursors in the literature: various aspects
of the contextual indexation of random variables and probabilities
of the kind shown are considered in Refs. \cite{larsson_kochen-specker_2002,dzhafarov_qualified_2014,Simon-Brukner-Zeilinger,Winter2014,svozil_how_2012,dzhafarov_all-possible-couplings_2013,Khr2005,Khr2008,Khr2009}.
The principal difference, however, is in the use of minimization of
$\Delta$ under the assumption that a jpd exists. This is a well-defined
mathematical problem, solvable in principle for any set of distributions
observed empirically. We will now compare and interrelate the two
approaches, NP and CbD, by applying them to the Leggett-Garg and the
EPR-Bell setups.

\section{Leggett-Garg }

Let us consider Leggett and Garg's $\pm1$-valued random variables,
$\mathbf{Q}_{1}$, $\mathbf{Q}_{2}$, and $\mathbf{Q}_{3}$ \cite{leggett_quantum_1985}.
Applying the NP approach, we seek signed probabilities $\mu$ for
$\left(\mathbf{Q}_{1},\mathbf{Q}_{2},\mathbf{Q}_{3}\right)$ that
are consistent with the observed correlations $\left\langle \mathbf{Q}_{i}\mathbf{Q}_{j}\right\rangle $
and individual expectations $\left\langle \mathbf{Q}_{i}\right\rangle $,
with the smallest possible value of the L1 norm $M\equiv\sum_{w}\left|\mu\left(w\right)\right|$,
where $w$ denotes all possible combinations of values $\left(q_{1},q_{2},q_{3}\right)$
for $\left(\mathbf{Q}_{1},\mathbf{Q}_{2},\mathbf{Q}_{3}\right)$.
Here, we use the standard notation $\left\langle \cdot\right\rangle $
for the expectation operator. This problem can be easily solved, as
we only have $2^{3}$ atomic elements $w$: $\left(1,1,1\right)$,
$\left(1,1,-1\right)$, ... , $\left(-1,-1,-1\right)$. Thus, for
$\mathbf{Q}_{1}$, $\mathbf{Q}_{2}$, and $\mathbf{Q}_{3}$, the minimal
L1 norm $1+\Gamma_{\min}$ satisfies 
\begin{eqnarray}
\Gamma_{\min} & =\max & \left\{ 0,-\frac{1}{2}+\frac{1}{2}S_{LG}\right\} ,\label{eq:Mm-LG}
\end{eqnarray}
where $S_{LG}$ is defined as 
\begin{equation}
S_{LG}\equiv\max_{\#^{-}=1,3}\{\pm\left\langle \mathbf{Q}_{1}\mathbf{Q}_{2}\right\rangle \pm\left\langle \mathbf{Q}_{1}\mathbf{Q}_{3}\right\rangle \pm\left\langle \mathbf{Q}_{2}\mathbf{Q}_{3}\right\rangle \},\label{eq:=00003D000023- in LG}
\end{equation}
where each $\pm$ in the expression should be replaced with $+$ or
$-$, and $\#^{-}$ indicates the possible numbers of minuses. Notice
that $S_{LG}\leq1$, which is equivalent to $\Gamma_{\min}=0$, is
a necessary and sufficient condition for the existence of a proper
jpd. 

Turning now to the CbD approach, we create a set of six random variables
\begin{equation}
\mathbf{Q}_{1,2},\mathbf{Q}_{1,3},\mathbf{Q}_{2,1},\mathbf{Q}_{2,3},\mathbf{Q}_{3,1},\mathbf{Q}_{3,2},\label{eq:singles}
\end{equation}
each indexed by the measurement conditions under which it is recorded:
for any two random variables recorded at moments $t_{i}$ and $t_{j}$,
with $i<j$, the $\mathbf{Q}_{i,j}$ designates the earlier variable
and $\mathbf{Q}_{j,i}$ the later one. We have thus three pairs of
variables with known jpds: 
\begin{equation}
\left(\mathbf{Q}_{1,2},\mathbf{Q}_{2,1}\right),\left(\mathbf{Q}_{1,3},\mathbf{Q}_{3,1}\right),\left(\mathbf{Q}_{2,3},\mathbf{Q}_{3,2}\right).\label{eq:pairs}
\end{equation}
A jpd can always be constructed for these pairs (e.g., they can always
be connected as stochastically independent pairs), but we seek a jpd
with the smallest value $\Delta_{\min}$ of 
\begin{equation}
\begin{array}{r}
\Delta=\Pr\left[\mathbf{Q}_{1,2}\neq\mathbf{Q}_{1,3}\right]+\Pr\left[\mathbf{Q}_{2,1}\neq\mathbf{Q}_{2,3}\right]+\Pr\left[\mathbf{Q}_{3,1}\neq\mathbf{Q}_{3,2}\right].\end{array}
\end{equation}
A classical joint exists for $\mathbf{Q}_{1}$, $\mathbf{Q}_{2}$,
and $\mathbf{Q}_{3}$ (no contextuality) if and only if a joint exists
for (\ref{eq:pairs}) with $\Delta=0$. The more we depart from the
classical joint, the larger the minimum value $\Delta_{\min}$. Thus,
$\Delta_{\min}$ can serve as a measure of contextuality. 

Requiring a jpd consistent with (\ref{eq:pairs}) means to assign
a probability to each of the $2^{6}$ possible values of these random
variables, 
\begin{equation}
\mathbf{Q}_{1,2}=\pm1,\mathbf{Q}_{1,3}=\pm1,\ldots,\mathbf{Q}_{3,2}=\pm1,
\end{equation}
constrained by being nonnegative and summing to the observed probabilities.
For instance, the probabilities assigned to all combinations with
$\mathbf{Q}_{1,2}=1$ and $\mathbf{Q}_{2,1}=-1$ should sum to the
observed $\Pr\left[\mathbf{Q}_{1,2}=1,\mathbf{Q}_{2,1}=-1\right]$.
A computer-assisted Fourier-Motzkin elimination algorithm gives the
following analytic expression for the minimum value of $\Delta$ consistent
with the observable pairs (\ref{eq:pairs}): 
\begin{equation}
\Delta_{\min}=\max\left\{ 0,-\frac{1}{2}+\frac{1}{2}S_{LG}\right\} .\label{eq:Delta-LG}
\end{equation}
This is a special case of the result in Ref. \cite{dzhafarov_generalizing_2014,DKL2014,KDL2014}.

Comparing the general expressions (\ref{eq:Mm-LG}) for $\Gamma_{\min}$
and (\ref{eq:Delta-LG}) for $\Delta_{\min}$ we see that the two
simply coincide: 
\begin{equation}
\Delta_{\min}=\Gamma_{\min}.
\end{equation}

\section{EPR-Bell}

We now turn to the EPR-Bell case where Alice and Bob have each two
distinct settings, $1$ and $2$, corresponding to four observable
random variables $\mathbf{A}_{1}$, $\mathbf{A}_{2}$, $\mathbf{B}_{1}$,
and $\mathbf{B}_{2}$. This notation implicitly contains the assumption
that the identity of Alice's measurements as random variables does
not depend on Bob's settings, and vice versa. It is well known \cite{fine_hidden_1982}
that under the no-signaling conditions the existence of the jpd is
equivalent to the CHSH inequalities being satisfied. Applying the
NP approach, the minimal L1 norm of the probability distribution is
given by \cite{oas_exploring_2014} 
\begin{equation}
\Gamma_{\min}=\max\left\{ 0,\frac{1}{2}S_{CHSH}-1\right\} ,\label{eq:M-inequality-AliceBob}
\end{equation}
where 
\begin{equation}
\begin{array}{r}
S_{CHSH}=\raisebox{0pt}[0pt][0pt]{\ensuremath{{\displaystyle \max_{\#^{-}=1,3}}}}\{\pm\left\langle \mathbf{A}_{1,1}\mathbf{B}_{1,1}\right\rangle \pm\left\langle \mathbf{A}_{1,2}\mathbf{B}_{1,2}\right\rangle \phantom{\mbox{\}}.}\\
\pm\left\langle \mathbf{A}_{2,1}\mathbf{B}_{2,1}\right\rangle \pm\left\langle \mathbf{A}_{2,2}\mathbf{B}_{2,2}\right\rangle \mbox{\}}.
\end{array}
\end{equation}
Here $\Gamma_{\min}=0$ corresponds to the CHSH inequalities, and
$\Gamma_{\min}>0$ to contextuality. 

Turning now to the CbD approach, we have four pairs of random variables,
\begin{equation}
\left(\mathbf{A}_{1,1},\mathbf{B}_{1,1}\right),\left(\mathbf{A}_{1,2},\mathbf{B}_{1,2}\right),\left(\mathbf{A}_{2,1},\mathbf{B}_{2,1}\right),\left(\mathbf{A}_{2,2},\mathbf{B}_{2,2}\right).\label{eq:AB pairs}
\end{equation}
Here, $\mathbf{A}_{i,j}$ denotes Alice's measurement under her setting
$i=1,2$ when Bob's setting is $j=1,2$, and analogously for $\mathbf{B}_{i,j}$.
We seek a jpd with the smallest value $\Delta_{\min}$ of 
\begin{equation}
\begin{array}{r}
\Pr\left[\mathbf{A}_{1,1}\neq\mathbf{A}_{1,2}\right]+\Pr\left[\mathbf{A}_{2,1}\neq\mathbf{A}_{2,2}\right]+\Pr\left[\mathbf{B}_{1,1}\neq\mathbf{B}_{2,1}\right]+\Pr\left[\mathbf{B}_{1,2}\neq\mathbf{B}_{2,2}\right].\end{array}
\end{equation}
No contextuality means $\Delta_{\min}=0$. A computer assisted Fourier-Motzkin
elimination algorithm yields (this is a special case of the result
in Ref. \cite{dzhafarov_generalizing_2014,DKL2014,KDL2014}) 
\begin{equation}
\Delta_{\min}=\max\left\{ 0,\frac{1}{2}S_{CHSH}-1\right\} .\label{eq:delta-AliceBob}
\end{equation}

We have the same simple coincidence the two measures as in the case
of the Leggett-Garg systems, 
\begin{equation}
\Delta_{\min}=\Gamma_{\min}.
\end{equation}

\section{Final remarks }

We have discussed two ways to measure contextuality. The direct approach,
named Contextuality-by-Default (CbD), assigns to each random variable
an index related to their context. If a system is noncontextual, a
jpd can be imposed on the random variables so that any two of them
representing the same property in different contexts always have the
same values. If the system is contextual, the minimum value of $\Delta$
in (\ref{eq:Delta-LG})-(\ref{eq:delta-AliceBob}) across all possible
jpds has the interpretation of how close a variable can be in two
different contexts: the larger the value the greater contextuality,
zero representing a necessary and sufficient condition for no contextuality. 

The other approach maintains the original set of random variables,
but requires negative (quasi-)probabilities. This leads to nonmonotonicity
(i.e., a set of outcomes can have a smaller probability than some
of its proper subsets), which is a characteristic of quantum interference.
The departure from a proper probability distribution is measured by
$\Gamma_{\min}$ in the minimum L1 norm $1+\Gamma_{\min}$. Similar
to the CbD approach, we use here a minimization principle that gives
the closest probability distribution to an ideal (but impossible)
jpd. The value of $\Gamma_{\min}$ has the interpretation of how contextual
the system is: a necessary and sufficient condition for no contextuality
is $\Gamma_{\min}=0$, and the larger the value of $\Gamma_{\min}$,
the more contextual the system is. 

As we have seen, in the case of EPR-Bell and Leggett-Garg systems
the two approaches lead to simple coincidence, $\Delta_{\min}=\Gamma_{\min}$.
The two measures, $\Gamma_{\min}$ and $\Delta_{\min}$, can be computed,
in principle, for any given system. For detailed examples of such
computations, see Appendix. 

Of the two measures of contextuality, $\Gamma_{\min}$ is computationally
much simpler, as it involves fewer random variables and a simpler
set of conditions (no nonnegativity constraints). However, CbD has
the advantage of being more general than NP, as it can include cases
where no NP distributions exist due to violations of the no-signaling
condition \cite{dzhafarov_probabilistic_2014,DKL2014,KDL2014}.

\paragraph{Acknowledgments. This work was supported by NSF grant SES-1155956
and AFOSR grant FA9550-14-1-0318. The authors are grateful to Samson
Abramsky, Guido Bacciagaluppi, Andrei Khrennikov, Jan-Åke Larsson,
and Patrick Suppes for helpful discussions. }

\appendix

\section{Proofs of statements}

\label{appendix:proofs}In this appendix, we describe how the analytic
results of the main text were obtained for each of the expressions
\eqref{eq:delta-AliceBob}, \eqref{eq:M-inequality-AliceBob}, \eqref{eq:Delta-LG},
and \eqref{eq:Mm-LG} of the main text.

\subsection{EPR-Bell: Contextuality-by-Default}

Following the computations of Dzhafarov and Kujala \cite[Text S3]{dzhafarov_all-possible-couplings_2013},
or the more general formulation in Ref. \cite{dzhafarov_generalizing_2014},
it can be shown that the observable distributions with probabilities
given by the matrices 

\begin{equation}
\begin{tabular}{c|cc|ccc|cc|ccccccccccccccccccccccccccccccccccccccccccccccccccccccccccccccccc}
 \cline{2-3} \cline{7-8}  &  \ensuremath{\mathbf{B}_{1,1}=+1} &  \ensuremath{\mathbf{B}_{1,1}=-1} &   &  \quad{}  &   &  \ensuremath{\mathbf{B}_{1,2}=+1} &  \ensuremath{\mathbf{B}_{1,2}=-1} &  \tabularnewline\cline{1-4} \cline{6-9} \multicolumn{1}{|c|}{\ensuremath{\mathbf{A}_{1,1}=+1}}  &  \ensuremath{p_{1,1}} &  \ensuremath{a_{1}-p_{1,1}} &  \multicolumn{1}{c|}{\ensuremath{a_{1}}}  &  \multicolumn{1}{c|}{}  &  \ensuremath{\mathbf{A}_{1,2}=+1} &  \ensuremath{p_{1,2}} &  \ensuremath{a_{1}-p_{1,2}} &  \multicolumn{1}{c|}{\ensuremath{a_{1}}}\tabularnewline\multicolumn{1}{|c|}{\ensuremath{\mathbf{A}_{1,1}=-1}}  &  \ensuremath{b_{1}-p_{1,1}} &  \ensuremath{1-a_{1}-b_{1}+p_{1,1}} &  \multicolumn{1}{c|}{\ensuremath{1-a_{1}}}  &  \multicolumn{1}{c|}{}  &  \ensuremath{\mathbf{A}_{1,2}=-1} &  \ensuremath{b_{2}-p_{1,2}} &  \ensuremath{1-a_{1}-b_{2}+p_{1,2}} &  \multicolumn{1}{c|}{\ensuremath{1-a_{1}}}\tabularnewline\cline{1-4} \cline{6-9}  &  \ensuremath{b_{1}} &  \ensuremath{1-b_{1}} &   &   &   &  \ensuremath{b_{2}} &  \ensuremath{1-b_{2}} &  \tabularnewline\cline{2-3} \cline{7-8} \multicolumn{1}{c}{}  &   &  \multicolumn{1}{c}{}  &   &   &  \multicolumn{1}{c}{}  &   &  \multicolumn{1}{c}{}  &  \tabularnewline\cline{2-3} \cline{7-8}  &  \ensuremath{\mathbf{B}_{2,1}=+1} &  \ensuremath{\mathbf{B}_{2,1}=-1} &   &   &   &  \ensuremath{\mathbf{B}_{2,2}=+1} &  \ensuremath{\mathbf{B}_{2,2}=-1} &  \tabularnewline\cline{1-4} \cline{6-9} \multicolumn{1}{|c|}{\ensuremath{\mathbf{A}_{2,1}=+1}}  &  \ensuremath{p_{2,1}} &  \ensuremath{a_{2}-p_{2,1}} &  \multicolumn{1}{c|}{\ensuremath{a_{2}}}  &  \multicolumn{1}{c|}{}  &  \ensuremath{\mathbf{A}_{2,2}=+1} &  \ensuremath{p_{2,2}} &  \ensuremath{a_{2}-p_{2,2}} &  \multicolumn{1}{c|}{\ensuremath{a_{2}}}\tabularnewline\multicolumn{1}{|c|}{\ensuremath{\mathbf{A}_{2,1}=-1}}  &  \ensuremath{b_{1}-p_{2,1}} &  \ensuremath{1-a_{2}-b_{1}+p_{2,1}} &  \multicolumn{1}{c|}{\ensuremath{1-a_{2}}}  &  \multicolumn{1}{c|}{}  &  \ensuremath{\mathbf{A}_{2,2}=-1} &  \ensuremath{b_{2}-p_{2,2}} &  \ensuremath{1-a_{2}-b_{2}+p_{2,2}} &  \multicolumn{1}{c|}{\ensuremath{1-a_{2}}}\tabularnewline\cline{1-4} \cline{6-9}  &  \ensuremath{b_{1}} &  \ensuremath{1-b_{1}} &   &   &   &  \ensuremath{b_{2}} &  \ensuremath{1-b_{2}} &  \tabularnewline\cline{2-3} \cline{7-8} \end{tabular}\label{eq:obs}
\end{equation}
are compatible with the connections
\begin{equation}
\begin{tabular}{c|cc|ccc|cc|cccccccccccccccccccccccccccccccccccccccccccccccccccccccccccccccccccccccccc}
 \cline{2-3} \cline{7-8}  &  \ensuremath{\mathbf{A}_{1,2}=+1} &  \ensuremath{\mathbf{A}{}_{1,2}=-1} &   &  \quad{}  &   &  \ensuremath{\mathbf{B}_{2,1}=+1} &  \ensuremath{\mathbf{B}_{2,1}=-1} &   &  \quad{}\quad{}\tabularnewline\cline{1-4} \cline{6-9} \multicolumn{1}{|c|}{\ensuremath{\mathbf{A}_{1,1}=+1}}  &  \ensuremath{a_{1}-\alpha_{1}} &  \ensuremath{\alpha_{1}} &  \multicolumn{1}{c|}{\ensuremath{a_{1}}}  &  \multicolumn{1}{c|}{}  &  \ensuremath{\mathbf{B}_{1,1}=+1} &  \ensuremath{b_{1}-\beta_{1}} &  \ensuremath{\beta_{1}} &  \multicolumn{1}{c|}{\ensuremath{b_{1}}}  &  \tabularnewline\multicolumn{1}{|c|}{\ensuremath{\mathbf{A}_{1,1}=-1}}  &  \ensuremath{\alpha_{1}} &  \ensuremath{1-a_{1}-\alpha_{1}} &  \multicolumn{1}{c|}{\ensuremath{1-a_{1}}}  &  \multicolumn{1}{c|}{}  &  \ensuremath{\mathbf{B}_{1,1}=-1} &  \ensuremath{\beta_{1}} &  \ensuremath{1-b_{1}-\beta_{1}} &  \multicolumn{1}{c|}{\ensuremath{1-b_{1}}}  &  \tabularnewline\cline{1-4} \cline{6-9}  &  \ensuremath{a_{1}} &  \ensuremath{1-a_{1}} &   &   &   &  \ensuremath{b_{1}} &  \ensuremath{1-b_{1}} &   &  \tabularnewline\cline{2-3} \cline{7-8} \multicolumn{1}{c}{}  &   &  \multicolumn{1}{c}{}  &   &   &  \multicolumn{1}{c}{}  &   &  \multicolumn{1}{c}{}  &   &  \tabularnewline\cline{2-3} \cline{7-8}  &  \ensuremath{\mathbf{A}_{2,2}=+1} &  \ensuremath{\mathbf{A}{}_{2,2}=-1} &   &   &   &  \ensuremath{\mathbf{B}_{2,2}=+1} &  \ensuremath{\mathbf{B}_{2,2}=-1} &   &  \tabularnewline\cline{1-4} \cline{6-9} \multicolumn{1}{|c|}{\ensuremath{\mathbf{A}_{2,1}=+1}}  &  \ensuremath{a_{2}-\alpha_{2}} &  \ensuremath{\alpha_{2}} &  \multicolumn{1}{c|}{\ensuremath{a_{2}}}  &  \multicolumn{1}{c|}{}  &  \ensuremath{\mathbf{B}_{1,2}=+1} &  \ensuremath{b_{2}-\beta_{2}} &  \ensuremath{\beta_{2}} &  \multicolumn{1}{c|}{\ensuremath{b_{2}}}  &  \tabularnewline\multicolumn{1}{|c|}{\ensuremath{\mathbf{A}_{2,1}=-1}}  &  \ensuremath{\alpha_{2}} &  \ensuremath{1-a_{2}-\alpha_{2}} &  \multicolumn{1}{c|}{\ensuremath{1-a_{2}}}  &  \multicolumn{1}{c|}{}  &  \ensuremath{\mathbf{B}_{1,2}=-1} &  \ensuremath{\beta_{2}} &  \ensuremath{1-b_{2}-\beta_{2}} &  \multicolumn{1}{c|}{\ensuremath{1-b_{2}}}  &  \tabularnewline\cline{1-4} \cline{6-9}  &  \ensuremath{a_{2}} &  \ensuremath{1-a_{2}} &   &   &   &  \ensuremath{b_{2}} &  \ensuremath{1-b_{2}} &   &  \tabularnewline\cline{2-3} \cline{7-8} \end{tabular}\label{eq:conn}
\end{equation}
if and only if
\begin{equation}
\begin{array}{l}
s_{0}\!\left(\left\langle \mathbf{A}_{1,1}\mathbf{B}_{1,1}\right\rangle \!,\left\langle \mathbf{A}_{1,2}\mathbf{B}_{1,2}\right\rangle \!,\left\langle \mathbf{A}_{2,1}\mathbf{B}_{2,1}\right\rangle \!,\left\langle \mathbf{A}_{2,2}\mathbf{B}_{2,2}\right\rangle \right)\\
\le6-s_{1}\!\left(\left\langle \mathbf{A}_{1,1}\mathbf{A}_{1,2}\right\rangle \!,\left\langle \mathbf{A}_{2,1}\mathbf{A}_{2,2}\right\rangle \!,\left\langle \mathbf{B}_{1,1}\mathbf{B}_{2,1}\right\rangle \!,\left\langle \mathbf{B}_{1,2}\mathbf{B}_{2,2}\right\rangle \right),
\end{array}\label{eq:CbD-s0}
\end{equation}
\begin{equation}
\begin{array}{l}
s_{1}\!\left(\left\langle \mathbf{A}_{1,1}\mathbf{B}_{1,1}\right\rangle \!,\left\langle \mathbf{A}_{1,2}\mathbf{B}_{1,2}\right\rangle \!,\left\langle \mathbf{A}_{2,1}\mathbf{B}_{2,1}\right\rangle \!,\left\langle \mathbf{A}_{2,2}\mathbf{B}_{2,2}\right\rangle \right)\\
\le6-s_{0}\!\left(\left\langle \mathbf{A}_{1,1}\mathbf{A}_{1,2}\right\rangle \!,\left\langle \mathbf{A}_{2,1}\mathbf{A}_{2,2}\right\rangle \!,\left\langle \mathbf{B}_{1,1}\mathbf{B}_{2,1}\right\rangle \!,\left\langle \mathbf{B}_{1,2}\mathbf{B}_{2,2}\right\rangle \right).
\end{array}\label{eq:CbD-s1}
\end{equation}
 where 
\begin{align*}
s_{0}(x_{1},\dots,x_{n}) & =\max\{\pm x_{1}\pm\dots\pm x_{n}:\text{even \# of \ensuremath{-}'s}\},\\
s_{1}(x_{1},\dots,x_{n}) & =\max\{\pm x_{1}\pm\dots\pm x_{n}:\text{odd \# of \ensuremath{-}'s}\},
\end{align*}
and where we use the parameterization by the 12 expectation variables
defined as
\begin{equation}
\left\langle \mathbf{A}_{i,j}\mathbf{B}_{i,j}\right\rangle =\left(4p_{ij}-1\right)-\left(2a_{i}-1\right)-\left(2b_{j}-1\right),\label{eq:ABcorr}
\end{equation}
\begin{equation}
\left\langle \mathbf{A}_{i,1}\mathbf{A}_{i,2}\right\rangle =1-4\alpha_{i}=1-2\Pr\left[\mathbf{A}_{i,1}\ne\mathbf{A}_{i,2}\right],\label{eq:Acorr}
\end{equation}
\begin{equation}
\left\langle \mathbf{B}_{1,j}\mathbf{B}_{2,j}\right\rangle =1-4\beta_{j}=1-2\Pr\left[\mathbf{B}_{1,j}\ne\mathbf{B}_{2,j}\right],\label{eq:Bcorr}
\end{equation}
\begin{equation}
\left\langle \mathbf{A}_{i}\right\rangle =2a_{i}-1,\label{eq:Amarginal}
\end{equation}
\begin{equation}
\left\langle \mathbf{B}_{j}\right\rangle =2b_{j}-1,\label{eq:Bmarginal}
\end{equation}
 for $i,j\in\text{\{1,2\}}.$

Writing the inequalities \eqref{eq:CbD-s0} and \eqref{eq:CbD-s1}
in terms of these expectations rather than in terms of probabilities
is the most economic way of presenting the 128 non-trivial inequalities
of the system, as the marginal probabilities $a_{1},a_{2},b_{1},b_{2}$
(or expectations $\left\langle \mathbf{A}_{1}\right\rangle ,\left\langle \mathbf{A}_{2}\right\rangle ,\left\langle \mathbf{B}_{1}\right\rangle ,\left\langle \mathbf{B}_{2}\right\rangle $)
vanish in this form. However, it should be noted that in addition
to these 128 inequalities, the form of the observed distributions
and connections itself imposes further 28 trivial constraints on the
12 expectation variables of the system: the probabilities within each
$2\times2$ matrix in \eqref{eq:obs} and \eqref{eq:conn} should
be nonnegative and sum to one. 16 of these trivial constraints pertain
to the observed distributions and 12 to the connections. In terms
of the expectations, these trivial constraints correspond to 
\begin{equation}
-1+|\left\langle \mathbf{A}\right\rangle +\left\langle \mathbf{B}\right\rangle |\le\left\langle \mathbf{A}\mathbf{B}\right\rangle \le1-|\left\langle \mathbf{A}\right\rangle -\left\langle \mathbf{B}\right\rangle |,\label{eq:trivial}
\end{equation}
for given marginals for each pair $(\mathbf{A},\mathbf{B})$ of random
variables in \eqref{eq:ABcorr}--\eqref{eq:Bcorr}. This expands to
four inequalities for each of the observed distributions and to three
inequalities for each of the connections (the two upper bounds in
\eqref{eq:trivial} coincide when $\left\langle \mathbf{A}\right\rangle =\left\langle \mathbf{B}\right\rangle $).
Although these trivial constraints can usually be assumed implicitly,
it is important to keep them explicitly in the system for the next
step.

Adding the equation 
\begin{align*}
\Delta= & \Pr\left[\mathbf{A}_{1,1}\ne\mathbf{A}_{1,2}\right]+\Pr\left[\mathbf{A}_{2,1}\ne\mathbf{A}_{2,2}\right]\\
 & +\Pr\left[\mathbf{B}_{1,1}\ne\mathbf{B}_{2,1}\right]+\Pr\left[\mathbf{B}_{1,2}\ne\mathbf{B}_{2,2}\right]\\
= & 2-\frac{1}{2}\left(\left\langle \mathbf{A}_{1,1}\mathbf{A}_{1,2}\right\rangle +\left\langle \mathbf{A}_{2,1}\mathbf{A}_{2,2}\right\rangle \right.\\
 & +\left.\left\langle \mathbf{B}_{1,1}\mathbf{B}_{2,1}\right\rangle +\left\langle \mathbf{B}_{1,2}\mathbf{B}_{2,2}\right\rangle \right)
\end{align*}
to the system and then eliminating the connection correlations $\left\langle \mathbf{A}_{1,1}\mathbf{A}_{1,2}\right\rangle $,
$\left\langle \mathbf{A}_{2,1}\mathbf{A}_{2,2}\right\rangle $, $\left\langle \mathbf{B}_{1,1}\mathbf{B}_{2,1}\right\rangle $,
$\left\langle \mathbf{B}_{1,2}\mathbf{B}_{2,2}\right\rangle $ from
the system using the Fourier--Motzkin elimination algorithm, we obtain
the system 
\begin{align}
 & -1+\frac{1}{2}S_{CHSH}\le\Delta\le4-\left[-1+\frac{1}{2}S_{CHSH}\right],\label{eq:delta1}\\
 & 0\le\Delta\le4-\left(\left|\left\langle \mathbf{A}_{1}\right\rangle \right|+\left|\left\langle \mathbf{A}_{2}\right\rangle \right|+\left|\left\langle \mathbf{B}_{1}\right\rangle \right|+\left|\left\langle \mathbf{B}_{2}\right\rangle \right|\right),\label{eq:delta2}
\end{align}
where we denote

\begin{equation}
\begin{array}{l}
S_{CHSH}=s_{1}\big(\left\langle \mathbf{A}_{1,1}\mathbf{B}_{1,1}\right\rangle ,\left\langle \mathbf{A}_{1,2}\mathbf{B}_{1,2}\right\rangle ,\\
\phantom{S_{CHSH}=s_{1}\big(}\left\langle \mathbf{A}_{2,1}\mathbf{B}_{2,1}\right\rangle ,\left\langle \mathbf{A}_{2,2}\mathbf{B}_{2,2}\right\rangle \big)
\end{array}
\end{equation}
as in the main text. This means that $\Delta$ is compatible with
the given observed probabilities if and only if the above inequalities
are satisfied. Since the set of possible values of $\Delta$ constrained
by (\ref{eq:delta1}) and (\ref{eq:delta2}) is known to be nonempty,
it follows that the minimum value of $\Delta$ is always given by
\[
\Delta_{\min}=\max\left\{ 0,\frac{1}{2}S_{CHSH}-1\right\} .
\]

\subsection{EPR-Bell: Negative probabilities}

The analogous result for the negative probabilities approach is that
the observable distributions \eqref{eq:obs} are obtained as the marginals
of some negative probability joint of $\mathbf{A}_{1}=\mathbf{A}_{1,1}=\mathbf{A}_{1,2},$
$\mathbf{A}_{2}=\mathbf{A}_{2,1}=\mathbf{A}_{2,2}$, $\mathbf{B}_{1}=\mathbf{B}_{1,1}=\mathbf{B}_{2,1},$
and $\mathbf{B}_{2}=\mathbf{B}_{1,2}=\mathbf{B}_{2,2}$ given by 
\begin{align*}
 & \Pr\left[\mathbf{A}_{1}=a'_{1},\mathbf{\, A}_{2}=a'_{2},\,\mathbf{B}_{1}=b'_{1},\,\mathbf{B}_{2}=b'_{2}\right]\\
 & \,=p^{+}(a'_{1},a'_{2},b'_{1},b'_{2})-p^{-}(a'_{1},a'_{2},b'_{1},b'_{2}),
\end{align*}
$a'_{1},a'_{2},b'_{1},b'_{2}\in\{1,-1\}$, for some nonnegative functions
$p^{+}$ and $p^{-}$ having a total probability mass value 
\[
M=\sum_{a'_{1},a'_{2},b'_{1},b'_{2}}p^{+}(a'_{1},a'_{2},b'_{1},b'_{2})+p^{-}(a'_{1},a'_{2},b'_{1},b'_{2})
\]
if and only if $M\ge1+\Gamma_{\min}$, where 
\[
\Gamma_{\min}=\max\left\{ 0,\frac{1}{2}S_{CHSH}-1\right\} .
\]

The computations are similar to those of the CbD approach, but there
are two general differences. First, in the CbD approach, the convex
range of the possible observed and connection expectations \eqref{eq:ABcorr}--\eqref{eq:Bmarginal}
over the convex polytope of all possible joints is obtained by looking
at these expectations at the $2^{8}$ vertices defining the polytope
of all joints and then applying a computer algorithm to find the set
of inequalities delineating the extreme values of the expectations
at these vertices. However, in the negative probabilities approach,
the joint is represented by the $2^{4}$ differences of the positive
and negative components of the distribution and so, although these
$2\cdot2^{4}$ components are nonnegative as in the CbD approach,
they do not need to sum to one. Hence, the joint is represented by
a convex cone rather than a bounded polytope. Still, a convex cone
is a special case of a general polytope and can be handled by the
same algorithms that we have used in the CbD approach.

Second, we do not need to apply the Fourier--Motzkin elimination algorithm
here as we have defined $M$ directly by the representation of the
joint so there are no extra variables we would need to eliminate.
This difference, however, is not really a difference between the two
approaches, as we could have done the same in the CbD approach as
well: we could have defined $\text{\ensuremath{\Delta}}$ directly
based on the joint of all eight variables without explicitly defining
the connection correlations \eqref{eq:Acorr}--\eqref{eq:Bcorr},
and then we would have obtained the result directly from the half-space
representation, as we do in the negative probabilities approach.

\subsection{Leggett--Garg: Contextuality-by-Default}

The results for Leggett--Garg $\mathbf{Q}_{1},\mathbf{Q}_{2},\mathbf{Q}_{3}$
can be obtained in the same way as for the EPR-Bell systems. In the
CbD approach, the observed correlations $\left\langle \mathbf{Q}_{1,2}\mathbf{Q}_{2,1}\right\rangle $,
$\left\langle \mathbf{Q}_{1,3}\mathbf{Q}_{3,1}\right\rangle $, $\left\langle \mathbf{Q}_{2,3}\mathbf{Q}_{3,2}\right\rangle $,
$\left\langle \mathbf{Q}_{1}\right\rangle =\left\langle \mathbf{Q}_{1,2}\right\rangle =\left\langle \mathbf{Q}_{1,3}\right\rangle $,
$\left\langle \mathbf{Q}_{2}\right\rangle =\left\langle \mathbf{Q}_{2,1}\right\rangle =\left\langle \mathbf{Q}_{3,2}\right\rangle $,
$\left\langle \mathbf{Q}_{3}\right\rangle =\left\langle \mathbf{Q}_{3,1}\right\rangle =\left\langle \mathbf{Q}_{3,2}\right\rangle $
are consistent with the connection correlations $\left\langle \mathbf{Q}_{1,2}\mathbf{Q}_{1,3}\right\rangle $,
$\left\langle \mathbf{Q}_{2,1}\mathbf{Q}_{2,3}\right\rangle $, $\left\langle \mathbf{Q}_{3,1}\mathbf{Q}_{3,2}\right\rangle $
if and only if these connection correlations are realizable with the
given marginals (i.e., each correlation $\left\langle \mathbf{AB}\right\rangle $
has to satisfy $-1+|\left\langle \mathbf{A}\right\rangle +\left\langle \mathbf{B}\right\rangle |\le\left\langle \mathbf{A}\mathbf{B}\right\rangle \le1-|\left\langle \mathbf{A}\right\rangle -\left\langle \mathbf{B}\right\rangle |$
as discussed in the EPR-Bell case above) and satisfy
\begin{equation}
\begin{array}{l}
s_{0}\left(\left\langle \mathbf{Q}_{1,2}\mathbf{Q}_{2,1}\right\rangle ,\left\langle \mathbf{Q}_{1,3}\mathbf{Q}_{3,1}\right\rangle ,\left\langle \mathbf{Q}_{2,3}\mathbf{Q}_{3,2}\right\rangle \right)\\
+s_{1}\left(\left\langle \mathbf{Q}_{1,2}\mathbf{Q}_{1,3}\right\rangle ,\left\langle \mathbf{Q}_{2,1}\mathbf{Q}_{2,3}\right\rangle ,\left\langle \mathbf{Q}_{3,1}\mathbf{Q}_{3,2}\right\rangle \right)\le4,
\end{array}
\end{equation}
\begin{equation}
\begin{array}{l}
s_{1}\left(\left\langle \mathbf{Q}_{1,2}\mathbf{Q}_{2,1}\right\rangle ,\left\langle \mathbf{Q}_{1,3}\mathbf{Q}_{3,1}\right\rangle ,\left\langle \mathbf{Q}_{2,3}\mathbf{Q}_{3,2}\right\rangle \right)\\
+s_{0}\left(\left\langle \mathbf{Q}_{1,2}\mathbf{Q}_{1,3}\right\rangle ,\left\langle \mathbf{Q}_{2,1}\mathbf{Q}_{2,3}\right\rangle ,\left\langle \mathbf{Q}_{3,1}\mathbf{Q}_{3,2}\right\rangle \right)\le4
\end{array}
\end{equation}
These two inequalities expand to 32 linear inequalities and there
are 21 trivial constraints.

Denoting 
\begin{align*}
\Delta & =\Pr\left[\mathbf{Q}_{1,2}\!\ne\!\mathbf{Q}_{1,3}\right]+\Pr\left[\mathbf{Q}_{2,1}\!\ne\!\mathbf{Q}_{2,3}\right]+\Pr\left[\mathbf{Q}_{3,1}\!\ne\!\mathbf{Q}_{3,2}\right]\\
 & =\frac{3}{2}-\frac{1}{2}\left(\left\langle \mathbf{Q}_{1,2}\mathbf{Q}_{1,3}\right\rangle +\left\langle \mathbf{Q}_{2,1}\mathbf{Q}_{2,3}\right\rangle +\left\langle \mathbf{Q}_{3,1}\mathbf{Q}_{3,2}\right\rangle \right)
\end{align*}
and eliminating the connection correlations from the system using
the Fourier--Motzkin algorithm, we obtain the system 
\begin{align}
-\frac{1}{2}+\frac{1}{2}S_{LG}\le\Delta & \le3-\left[-\frac{1}{2}+\frac{1}{2}S_{LG}^{0}\right],\\
0\le\Delta & \le3-\left|\left\langle \mathbf{Q}_{1}\right\rangle \right|-\left|\left\langle \mathbf{Q}_{2}\right\rangle \right|-\left|\left\langle \mathbf{Q}_{3}\right\rangle \right|,
\end{align}
where we denote 
\begin{align*}
S_{LG} & =s_{1}\left(\left\langle \mathbf{Q}_{1,2}\mathbf{Q}_{2,1}\right\rangle ,\left\langle \mathbf{Q}_{1,3}\mathbf{Q}_{3,1}\right\rangle ,\left\langle \mathbf{Q}_{2,3}\mathbf{Q}_{3,2}\right\rangle \right),\\
S_{LG}^{0} & =s_{0}\left(\left\langle \mathbf{Q}_{1,2}\mathbf{Q}_{2,1}\right\rangle ,\left\langle \mathbf{Q}_{1,3}\mathbf{Q}_{3,1}\right\rangle ,\left\langle \mathbf{Q}_{2,3}\mathbf{Q}_{3,2}\right\rangle \right).
\end{align*}
That is, $\Delta$ is consistent with the observed probabilities if
and only if the above inequalities are satisfied. It follows that
the minimum value of $\Delta$ is given by 
\[
\Delta_{\min}=\max\left\{ 0,\frac{1}{2}S_{LG}-\frac{1}{2}\right\} .
\]

\subsection{Leggett--Garg: Negative probabilities}

With the same additional comments as in the negative probability calculations
for the EPR-Bell case, our calculations show that the observable probabilities
$\Pr\left[\mathbf{Q}_{12}=q_{1},\mathbf{Q}_{21}=q_{2}\right]$, $\Pr\left[\mathbf{Q}_{13}=q_{1},\mathbf{Q}_{31}=q_{3}\right]$,
$\Pr\left[\mathbf{Q}_{23}=q_{2},\mathbf{Q}_{32}=q_{3}\right]$, $q_{1},q_{2},q_{3}\in\{0,1\},$
can be obtained as the marginals of a negative probability jpd of
$\mathbf{Q}_{1}=\mathbf{Q}_{1,2}=\mathbf{Q}_{1,3}$, $\mathbf{Q}_{2}=\mathbf{Q}_{2,1}=\mathbf{Q}_{3,2}$,
and $\mathbf{Q}_{3}=\mathbf{Q}_{3,1}=\mathbf{Q}_{3,2}$ with the total
probability mass of $M$ if and only if $M\ge1+\Gamma_{\min}$, where
\[
\Gamma_{\min}=\max\left\{ 0,\frac{1}{2}S_{LG}-\frac{1}{2}\right\} .
\]

\end{document}